# New Insight into Electronic Shells of Metal Clusters: Analogues of Simple Molecules

Longjiu Cheng (程龙玖),[1,a)] and Jinlong Yang (杨金龙)[2,a)]

[1]*School of Chemistry & Chemical Engineering, Anhui University, Hefei, Anhui, 230039, People's Republic of China.*

[2]*Hefei National Laboratory for Physics Sciences at Microscale, University of Science & Technology of China, Hefei, Anhui, 230026, People's Republic of China.*

[a)] Corresponding authors. E-mail: clj@ustc.edu; jlyang@ustc.edu.cn.

Using Li clusters, a prototype of simple metals, as a test case, we theoretically find that metal clusters can mimic the behavior of simple molecules in electronic shells. It is found that $Li_{14}$, $Li_{10}$, and $Li_8$ clusters are exact analogues of $F_2$, $N_2$, and $CH_4$ molecules.

Electronic shell closing has successfully explained the observed experimental abundances of alkali-metal clusters [1-5]. In this framework, a cluster is modeled as a superatom, in which valence electrons are delocalized in the cluster volume and fill discrete energy levels. There are several degrees of sophistication of this model, among which the spherical jellium model [1,6] is widely accepted. The spherical jellium model assumes a uniform background of positive charge of the cluster's atomic nuclei and their innermost electrons, in which valence electrons move and are subjected to an external potential. The infinitely deep spherical well and the harmonic well are the simplest forms of the potential, in which the former gives the series of magic numbers of 2, 8, 18, 20, 34, 40, 58, … , and the later gives the series of 2, 8, 20, 40, 70, … . The experimental spectra of sodium clusters [1] reveal high peaks at 8, 20, 40 in agreement with both models. The appropriate aufbau rule of delocalized "superatomic orbitals" of sodium clusters is | $1S^2$ | $1P^6$ | $1D^{10}$ $2S^2$ | $1F^{14}$ $2P^6$ | $1G^{18}$ $2D^{10}$ $3S^2$ | …, wherein S–P–D–F–G– denote the angular momentum quantum number and the 1–2–3– correspond to the radial nodes [2]. However, exceptional stability is associated with a total count of 2, 8, 18, 34, 58, … for ligand-protected gold clusters [7] in agreement with only the spherical model. The smaller peaks in the experimental spectra of sodium clusters [1], however, e.g., $N$=10, 12, 14, and 26 cannot all be understood in the framework of strictly spherical jellium model, where the cluster is far from spherical. To explain the smaller peaks, Clemenger [8] suggested an ellipsoidal shell model.

In this letter we show a more generalized insight into the non-spherical shells. Rare gas atoms have filled electronic shells, and are chemically very inert. The other elements are active in atomic state but can compose stable molecules, in which each atom has filled electronic shells by sharing valence pairs, which is just the basic idea of valence bond (VB) theory. Magic numbers of superatoms ($N$=8, 20, 40, …) can mimic the behavior of rare gas atoms in electronic shells. We found that the smaller peaks in mass spectra [1] may correspond to the superatomic molecules, i.e., superatoms can compose superatomic molecules by sharing valence pairs.

Lithium is the first alkali-metal, which can be considered as an ideal prototype for simple metals. In cluster physics, lithium cluster is a suitable model system for studying many fundamental physical properties of simple metal clusters. Thus, in this letter we also select lithium cluster to verify the concept of superatomic molecule. Geometries of small Li clusters have been subsequently studied by various density functional theory (DFT) and *ab initio* methods [9-13]. The energy landscape of Li clusters is rather smooth, so we revisit the global minimum (GM) structures of Li$_N$ clusters with $N = 2 - 26$ by unbiased global search of the DFT landscape using the genetic algorithm implemented in our group. There are certainly gaps in relative energies of different isomers between different theoretical methods. However, this letter focuses only on the most stable magic numbers where different methods should get consistent results, so we just simply choose the famous hybrid functional known as B3LYP without verification with the all-electron 6-311+G* basis set. All DFT calculations are carried out on Gaussian 09 package [14], and molecular visualization is performed using MOLEKEL 5.4 [15].

The energies of the resulting putative GMs of Li$_N$ clusters at B3LYP/6-311+G* with $N$ = 2 - 26 are depicted in Fig. 1 in a manner that emphasizes particular stable minima or "magic numbers" simulating the experimental mass spectra. Our simulated mass spectra are in good agreement with the experimental spectra [1] at both large ($N$ = 8, 20) and small ($N$ = 10, 12, 14) peaks at $N \leq 20$ indicating high reliability of our computational results. Bond length in the dimmer (Li$_2$) is 2.71 angstrom; Li$_4$ ($D_{2h}$) is a planar motif in all-metal σ-antiaromaticity [16]; Li$_7$ is a ideal tetrahedron in $D_{5h}$ symmetry; Li$_8$ ($T_d$) is a tetrahedron with all four faces capped; Li$_{10}$ ($D_{2d}$) is a prolate body-fused bi-tetrahedron sharing four nuclei; Li$_{12}$ ($C_{2v}$) and Li$_{14}$ ($D_{4h}$) are both prolate clusters; Li$_{19}$ ($C_{2v}$) and Li$_{20}$ ($C_s$) are in same packing style (all in Fig. 1). Stability of small Li clusters is mainly determined by the electronic shells (e.g., Li$_8$ and Li$_{20}$), and contributions of the atomic shells (e.g., Li$_7$ and Li$_{19}$) are also very clear. The filled electronic shells of Li$_{20}$ have been fully discussed using the jellium model [1-3], and we will tell a new story about the odd-even effect and the geometric and electronic structures of Li$_8$, Li$_{10}$, Li$_{12}$ and Li$_{14}$.

Firstly, we focus on the prolate double-core Li$_{14}$. Based on our concept of superatomic molecule, the prolate cluster can be seen as an union of two ten-center seven-electron (10c-7e) spherical superatoms sharing a six-nucleus octahedron (see Fig. 2a). The 10-7e superatom mimics the behavior of F atom in valence electronic shells ($s^2p^5$), and can be seen as a super F atom. The canonical Kohn-Shan molecular orbital (MO) diagrams (Fig. 3a) reveal clearly that Li$_{14}$ mimics the behavior of F$_2$ in electronic structures [17]. The first two valence MOs (HOMO-4, and HOMO-3) of $Li_{14}$ are super $\sigma_s$ and $\sigma^*_s$; HOMO-2 is a doubly degenerate π-bond MOs ($\pi_{px}$ and $\pi_{py}$); HOMO-1 is a super $\sigma_{pz}$; HOMO is a doubly degenerate π-antibond MOs ($\pi^*_{px}$ and $\pi^*_{py}$); LUMO+3 is a super $\sigma^*_{pz}$. Two δ-type MOs (LUMO and LUMO+1) and a $\sigma_{2s}$-type MO (LUMO+2) are lower than the $\sigma^*_{pz}$ MO. The reason may be that the shoulder-by-shoulder MOs (π and δ) are favored more at a shorter atom-atom/superatom-superatom distance, which is also the reason for the difference of the order of $\pi_{px,py}$ and $\sigma_{pz}$ between $Li_{14}$ and $F_2$.

The canonical MO pictures reveal that $Li_{14}$ is an analogue of $F_2$ based on the MO theory. Next, we will compare the bonding pictures based on the VB theory. To give a straightforward view of superatom-superatom bond, we selected the Adaptive Natural Density Partitioning (AdNDP) method as a tool for chemical bonding analysis. This method was recently developed by Zubarev and Boldyrev [18] and successfully used to analyze chemical bonding in organic molecules and clusters [18-23]. AdNDP recovers both Lewis bonding elements and delocalized bonding elements (*n*c-2e), which is a method of description of the chemical bonding combining the compactness and intuitive simplicity of Lewis theory with the flexibility and generality of canonical MO theory. As shown in Fig. 3b, AdNDP analysis reveals three 10c-2e super lone-pairs (LPs) in each superatom with occupancy number ON = 1.99-2.00 |e| in $Li_{14}$ (*s*, $p_x$, and $p_y$), and one 14c-2e super σ-bond (ON = 2.00 |e|). The bonding framework of $Li_{14}$ is an exact analogue of that of $F_2$ (Fig. 3c) based on the VB theory.

Secondly, we investigate the chemical bonding of $Li_{10}$. Similarly, the prolate cluster can be seen as an union of two 7c-5e spherical superatoms (super N), in which a four-nucleus tetrahedron and there covalent pairs (triple bond) are shared by the two super N (see Fig. 2b). The canonical MO diagrams (Fig. 4a) reveal that $Li_{10}$ is a complete analogue of $N_2$ in both MO shapes and orders, except for a $\sigma_{2s}$-type MO below $\sigma^*_{pz}$ MO. AdNDP analysis reveals one 5c-2e super $s$-type LP (ON = 1.96 |e|) in each superatom, one 10c-2e super $\sigma$-bond (ON = 2.00 |e|), and two 10c-2e super $\pi$-bonds (ON = 2.00 |e|) in $Li_{10}$ (Fig. 4b), which is same with $N_2$ (Fig. 4c) in bonding framework.

The situation in $Li_{12}$ is also similar. As shown in Fig. 2c, the prolate cluster can be seen as an union of a 7c-5e spherical superatoms (super N) and a 10c-7e spherical superatoms (super F), in which a five-nucleus bi-tetrahedron and two covalent pairs (one $\sigma$ and one $\pi$, double bond) is shared by this two superatoms. Thus, $Li_{12}$ is an analogue of NF molecule.

Lastly, we tell a new story of $Li_8$ other than the spherical jellium model. It is found that the peak at $N = 8$ is much smaller than the peak at $N = 20$ in the experimental spectra [1]. Moreover, the tetrahedral $Li_8$ is far from spherical. Thus, we think $Li_8$ is also a superatomic molecule instead of super Ne atom due to the size effect in atomic shells. After comparisons of geometries, canonical MO diagrams, and chemical bonding framework of AdNDP localization between $Li_8$ and $CH_4$, we conclude that, without any hesitation, $Li_8$ is a super $CH_4$ molecule (see Fig. 5). The central tetrahedron is a 4c-4e super C and the four capped tetrahedra are 4c-1e super H. A three-nucleus triangle and one covalent pair are shared by the super σ-CH.

After the perfect analogy of Li clusters, a prototype of simple metals, with simple molecules, now we can give conclusions on our super VB model: (i) metal clusters can mimic the behavior of simple molecules in electronic shells under both the MO and VB theories; (ii) the border between bonded superatoms is not so clear as between atoms, in which nuclei at the border may be shared by several superatoms.

Actually, chemical bond is just a definition of certain behavior of electrons, however, has achieved great success in explanation and prediction for the properties and structures of molecules and materials. Perhaps, our super VB theory is telling the same story with Clemenger's ellipsoidal shell model [8] from different ways. Although there is no unfathomable theory, our simple super VB model will be potentially very useful in the areas of metal clusters, nanoalloys, and materials, in which super covalent bonds may also exist as in Li clusters. Superatom theory has extended the periodic table to three dimensions [24-26], and the super VB model will open a new area of superatom-based clusters and materials.

It is a pleasure to thank Professor Boldyrev for the AdNDP program. This work is financed by the National Natural Science Foundation of China (20903001, 21121003, 91021004), by the 211 Project and the outstanding youth foundation of Anhui University, and by the National Key Basic Research Program of China (2011CB921404).

[a] Corresponding authors. Email: clj@ustc.edu; jlyang@ustc.edu.cn.

bonding features. Thus, as a model study, for clarity and simplicity, a small 3-21G basis set is used for MO and chemical bonding analysis in this letter.

**Figure captions**

**FIG. 1.** (color on line) Energies of the GMs of Li$_N$ clusters at B3LYP/6-311+G* relative to $E_{ave}$, a fit to the energies of the GMs at size ratio $2 \leq N \leq 26$ using the form $a + bN^{1/3} + cN^{2/3} + dN$, which represents the average energy of the GMs. Upward peaks, as labeled, represent particular stable minima or "magic numbers" relative to the "average energy" of the GMs. The structures and point groups of the upward peaks are labeled in the figure, where Li atoms may be given in different colors for better viewing.

**FIG. 2.** (color on line). Di-superatomic model for the prolate (a) Li$_{14}$, (b) Li$_{10}$, and (c) Li$_{12}$ clusters.

**FIG. 3.** (color on line). (a) Comparison of the geometries and Kohn-Sham MO diagrams of Li$_{14}$ cluster and F$_2$ molecules. (b) AdNDP localized natural bonding orbitals of Li$_{14}$ cluster. (c) AdNDP localized natural bonding orbitals of F$_2$ molecule.

**FIG. 4.** (color on line). (a) Comparison of the geometries and Kohn-Sham MO diagrams of Li$_{10}$ cluster and N$_2$ molecules. (b) AdNDP localized natural bonding orbitals of Li$_{10}$ cluster. (c) AdNDP localized natural bonding orbitals of N$_2$ molecule.

**FIG. 5.** (color on line). Comparison of the (a) geometries, (b) Kohn-Sham MO diagrams, and (c) AdNDP localized natural bonding orbitals of Li$_8$ cluster and CH$_4$ molecule.

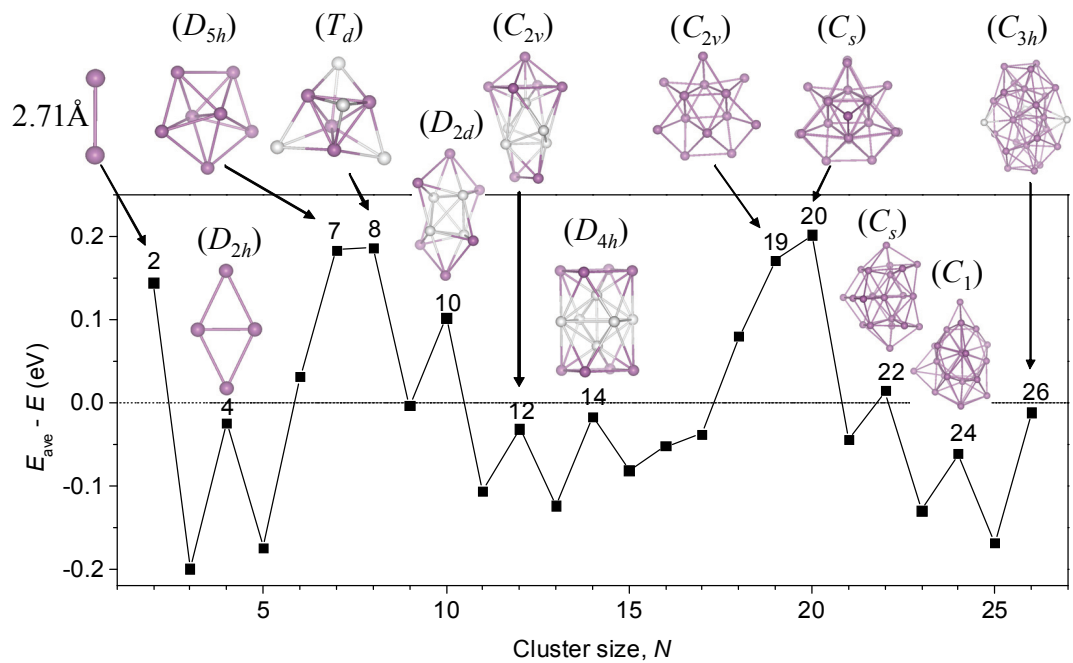

FIG. 1

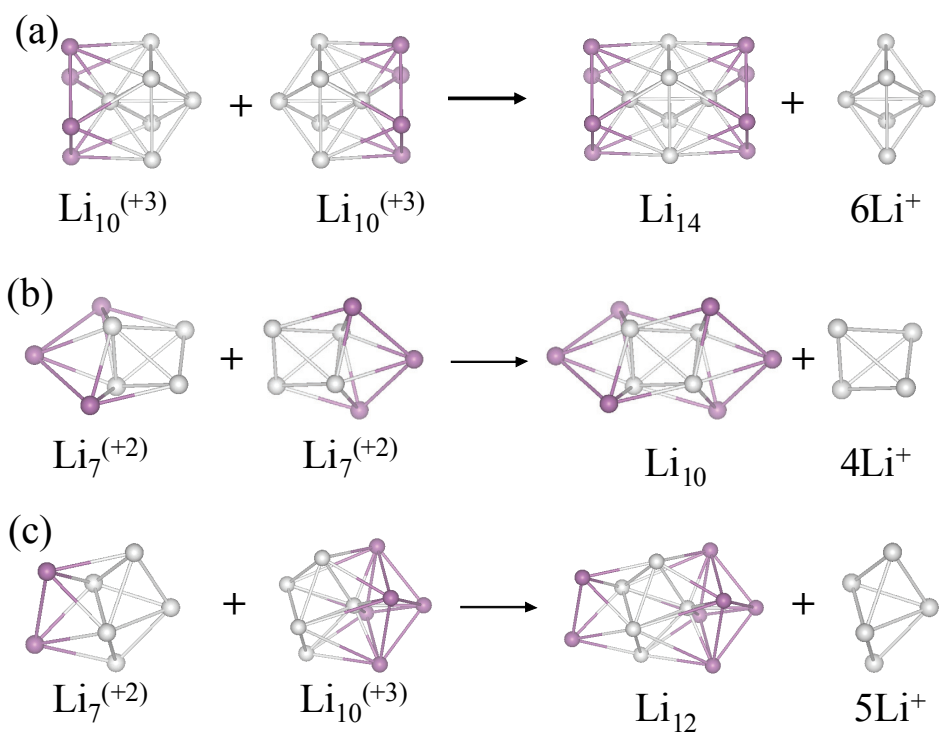

**FIG. 2**

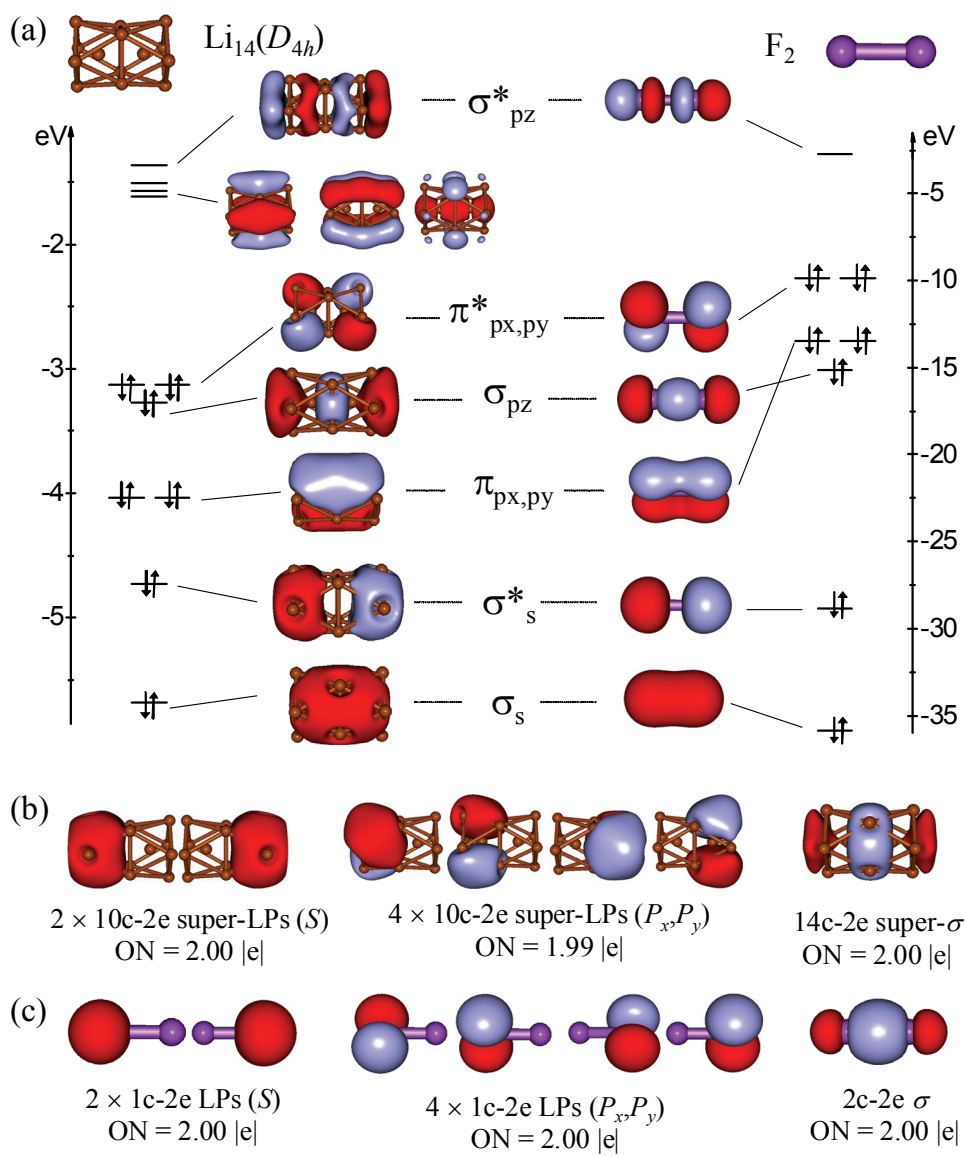

**FIG. 3**

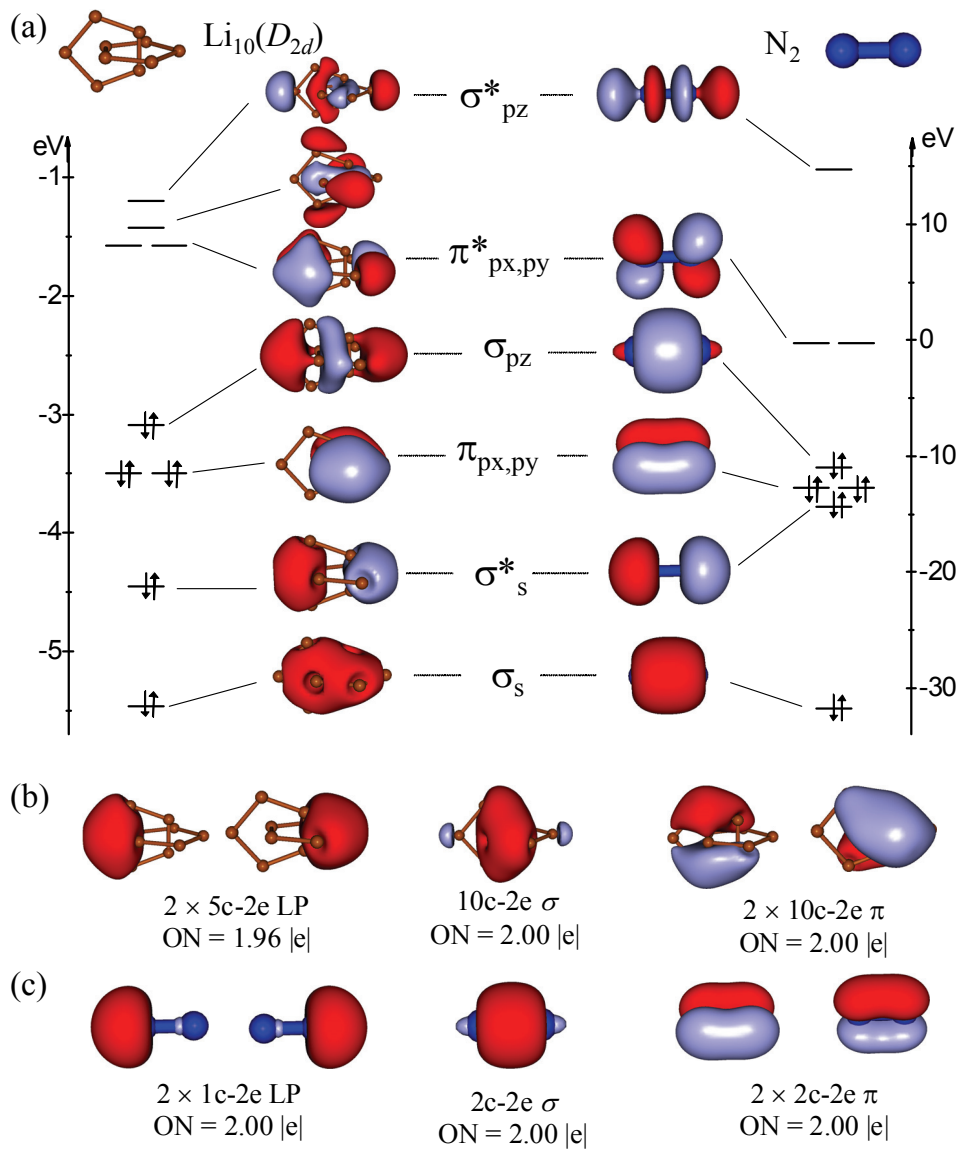

**Fig. 4**

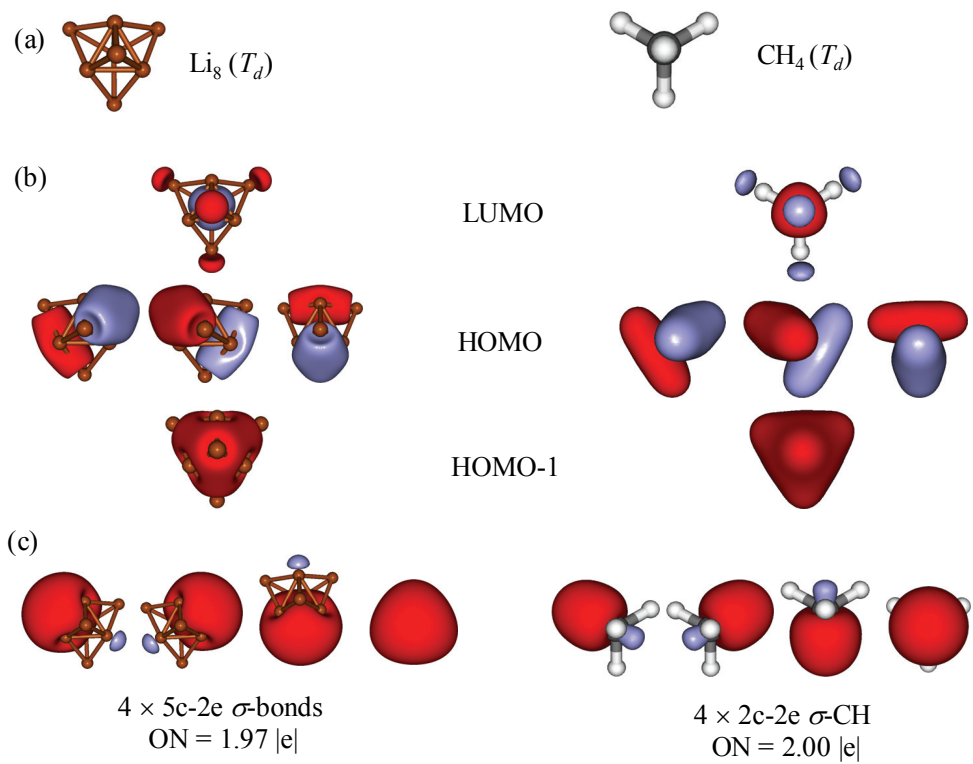

Fig. 5